# Study of Uniform Jitter Mechanism for Metric-based Wireless Routing

Shahbaz Rezaei
Department of Computer Engineering
Sharif University of Technology
Tehran, Iran
shrezaei@ce.sharif.edu

Ali Mohammad Afshin Hemmatyar
Department of Computer Engineering
Sharif University of Technology
Tehran, Iran
hemmatyar@sharif.edu

*Abstract*— Many wireless protocols wait a small and random amount of time which is called jitter before sending a packet to avoid high contention and packet collision. Jitter has been already proposed for many routing protocols including AODV and LOADng. However, since they do not consider any link quality parameters or metrics (such as ETX) in routing, they fail to be efficient in metric-based routing protocols. A metric-based jitter mechanism is proposed in this paper and a closed form expression is derived that enables us to obtain probability of delay inversion for all jitter mechanisms available. Simulation results are also presented to show performance of different jitter mechanisms.

*Keywords—Jitter mechanism, reactive routing protocol, wireless network.*

### Introduction

Wireless network is prone to high contention and collision due to it shared nature. In many cases, for instance when an event occurs in some region of wireless sensor network, a large number of nodes embark on generating and sending packets informing that event simultaneously. This common behavior contributes to high collision which degrades the throughput of the network dramatically. Hence, it could be reasonable to prevent nodes from sending their packet immediately. In many protocols, the idea of waiting for a small, random amount of time is exploited to alleviate collision. For many routing protocols such as AODV [6] and LOADng [7], a random value of delay, called jitter, has been already recommended in route discovery stage [1].

Simple jitter mechanism in which jitter is selected by a uniform random variable is analyzed mathematically in [2]. It is shown that the standard jitter mechanism is not efficient in many cases. Therefore, two other mechanisms, called window and adaptive jitter, have been proposed to solve some deficiencies of simple jitter mechanism [3]. Window jitter mechanism has been proposed to reduce the probability of delay inversion, a phenomenon in in which RREQ packet in the longer route (or worse route regarding any metric) reaches the destination sooner than the better route. In order to consider link metric or quality in jitter mechanism, adaptive jitter mechanism has been proposed. In [4,5], a closed form expression is obtained that gives the probability of delay inversion for window jitter mechanism. However, there is no mathematical expression to obtain probability of delay inversion in adaptive jitter mechanism or any arbitrary jitter mechanism.

Although jitter mechanisms are proposed to reduce collision, it can be exploited in routing protocols to find better routes. Adaptive jitter mechanism uses link metric (such as ETX [8]) to obtain a random jitter to exploit this feature [3].

Authors in [10] have also extensively studied the impact of using different random distribution on routing metrics. However, they did not obtain a closed-form expression to obtain the probability of delay inversion of a whole path.

In this paper, we derive a closed form expression that enables us to obtain probability of delay inversion of uniform distribution, using any arbitrary shaping function, as defined in [10]. We also illustrate results of our simulation of our jitter mechanism that can dramatically improve routes found in route discovery stage.

## I. BACKGOUND

### A. On-demand Routing Protocols

Routing protocols which try to find a route towards destination only when it is required are called on-demand routing protocols. One of the most important phases of these routing protocols is discovery stage in which the source sends and floods a route request packet (RREQ) over the network to find the destination. When the RREQ reaches the destination, a route is being formed and announced to the source by route reply packet. Hence, the way these RREP packets are forwarded has a dramatic influence on the quality of route formed in this stage. When the route failure occurs, the other phase, called route maintenance, starts to recover the route or find a new route.

### B. Jittering in Route Discovery Stage

If all nodes try to send and flood RREP packet simultaneously, most packets will be dropped due to high collision. To solve this issue, nodes are recommended to send the RREQ packet after a random amount of delay in RFC 5148 [1]. This essentially reduces the collision. Additionally, by imposing different value of delay, a routing protocol can find more desirable routes. For instance, nodes with low metric can be forced to send their RREQ with high value of delay to increase the chance of having a route with high metric. That is the reason why jittering mechanism are so important.

## II. DELAY INVERSION AND JITTERING METHODS

In this section, the delay inversion effect, an undesirable repercussion of jitter mechanisms, is introduced. Then, window and adaptive jittering designed for hop-based routing and metric-based routing, respectively, are introduced. Finally, a better adaptive jitter mechanism for tackling the delay inversion in metric-based routing as well as a formula to obtain probability of delay inversion is provided.

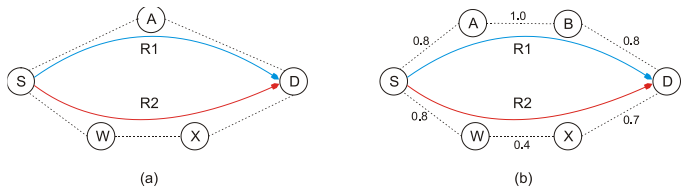

Fig. 1 Two possible routes for different routing mechanism: (a) when routing metric is hop count, the shortest path, R1, is optimal. (b) When a metric is assigned for each link, the best route would be the path with highest average metric, namely R1.

*A. The Delay Inversion*

In Fig.1 (a), there are two routes available from the source, S, to destination, D. Assuming that the hop count is used as a metric, the shortest (or better) route would be R1. Using random jitter in route discovery stage, however, a routing protocol may choose the worse path which is R2. In other words, the RREQ packet may encounter more delay at the node A than the nodes W and X together. This undesirable phenomenon is called delay inversion effect [4].

This undesirable effect can also occur in networks where there is a routing metric rather than hop count (e.g queue length, node's energy, etc). Assume that the metric shown in Fig.1 (b) represents available bandwidth. In this case, the route R1 obviously has more bandwidth than R2. However, if a routing protocol uses a simple random jitter without considering links' metric; both routes would have the same chance of getting selected since they have the same length (hop count). Therefore, it is of paramount importance to use a jitter mechanism wisely to be able to choose better routes.

The delay inversion probability can be obtained mathematically in different cases and scenarios. In the following sections, we will introduce several jittering techniques and mathematical formula to obtain their delay inversion so as to compare them.

*B. Hop-based Jitter*

To mitigate the problem of collision and simultaneous transmission, RFC 5148 [1] recommends a jitter-based transmission in which nodes delay each transmission by a random value from a uniform distribution $U \sim [0, J_{max}]$. $J_m$ is the maximum value of delay which is considered constant for a whole network. Although this mechanism can effectively reduce number of collisions, it slows the route discovery process and increases the probability of delay inversion [4].

In hop-based routing metric, where the shortest path is more preferable, a deterministic delay can completely eliminate the probability of delay inversion. In the deterministic approach, each node sends packets after $J_{max}$ milliseconds and, as a result, there is no randomness. Needless to say, this approach has the slowest route discovery stage as well as high collision probability [3].

Window jitter mechanism is proposed in [5] which is a trade-off between probability of delay inversion and route discovery time. In the window jitter mechanism, each node delays its transmissions by a random value from a uniform distribution $U \sim [\alpha J_{max}, J_{max}]$ that $\alpha = 0$ is tantamount to the method of RFC 5148 and $\alpha = 1$ is tantamount to the deterministic approach.

*C. Metric-based Jitter*

Although a few jittering methods are proposed for hop-based routing protocols, in which the shortest path is more preferable, jitter mechanisms for metric-based routing protocols need further scrutiny. Given a link metric $M \in (0,1)$ (m=1 indicates high quality links), authors in [5] proposed adaptive jitter mechanism which selects jitter values uniformly within $[(1-m)J_{max}, J_{max}]$ough the average delay imposed on better links would be lower, the probability of delay inversion is not insignificant owing to the fix upper bound.

To reduce the probability of delay inversion as well as route discovery time of better routes [10], delay values can be obtained from a random variable $U_C \sim [(1-m)J_{max}, (1-m)J_{max} + C]$, where C indicates the range of $U_c$, a trade-off between number of collisions and discovery time. The smaller the value of C is, the higher the number of collisions would be. The bigger the value of C is, the longer the process of route discovery would be. Since the distribution of high quality and low quality links has less overlap in general, the probability of delay inversion is lower than the adaptive jitter mechanism in [5].

*D. Analysis*

Let $\Omega_1, \Omega_2, \dots, \Omega_n$ be independent uniform random variables on $(a_i, b_i)$ and also $b_i > a_i$. Moreover, let $l_i = b_i - a_i$, $A_n = \sum_{i=1}^{n} a_i$ and $B_n = \sum_{i=1}^{n} b_i$. Now, S, a set containing all possible sum of i arbitrary elements of $\{l_1, l_2, \dots, l_n\}$, is defined as follows [9]

$$S = \{\omega_1 l_1 + \omega_2 l_2 + \cdots + \omega_n l_n \mid \omega_i \in \{0,1\}, i = 1, 2, \dots n\} \quad (1)$$

Having been sorted in ascending order, $S = \{s_1, s_2, \dots, s_{2^n}\}$ is used to define the function, $r_n(x)$, as follows

$$r_n(x) = max\{j \mid x - A_n - s_j > 0 \text{ and } s_j \in S \text{ for } j = 1, \dots, 2^n\} \quad (2)$$

In other words, $r_n(x)$ indicates the maximum of index j in which $A_n + s_j < x$. Assuming $W_n$ is a subspace of a n-dimensional Euclidean vector space $\{0,1\}^n$, for each j ($j = 1, 2, \dots, 2^n$), there exists a vector $w_{nj} = (\Omega_1, \Omega_2, \dots, \Omega_n)$ such that

$$s_j = \Omega_1 l_1 + \Omega_2 l_2 + \dots + \Omega_n l_n \quad (3)$$

Then, density function of sum of n independent random variable, uniformly distributed on $(a_i, b_i)$, is [9]

$$h_n(x) = \frac{\sum_{j=1}^{r_n(x)} (-1)^{\|w_{nj}\|} (x - A_n - s_j)^{n-1} I_{(A_n, B_n)}(x)}{(n-1)! \prod_{i=1}^{n} (b_i - a_i)} \quad (4)$$

where $\|.\|$ is the norm of the vector and $I_{(A_n,B_n)}(x)$ is the indicator function. Hence, the distribution function can be obtained as follows

$$H_n(x) = \frac{\sum_{j=1}^{r_n(x)} (-1)^{\|w_{nj}\|} (x - A_n - s_j)^n I_{(A_n,B_n)}(x)}{(n)! \prod_{i=1}^{n}(b_i - a_i)} + I_{(B_n,\infty)}(x) \quad (5)$$

The following theorem can be used to obtain the probability of delay inversion of any jitter mechanism which uses uniform distribution.

**Theorem 1:** Let $R_1$ and $R_2$ be two routes and $R_1$ be better than $R_2$ regarding an arbitrary routing metric. Using uniform distribution in jitter mechanism, the probability of delay inversion is

$$P(R_1 > R_2) = \xi_n \bar{\xi}_m \sum_{j=1}^{2^n} \sum_{k=1}^{2^m} (-1)^{\|w_{nj}\| + \|\bar{w}_{mk}\|}$$
$$\times \left[ {}_2F_1\left(m+1, 1-n; m+2; \frac{x - \bar{A}_m - \bar{s}_k}{A_n + s_j - \bar{A}_m - \bar{s}_k}\right) \right.$$
$$\left. \times \frac{(\bar{A}_m + \bar{s}_k - A_n - s_j)^{n-1}}{(m+1) \times (x - \bar{A}_m - \bar{s}_k)^{-(m+1)}} \right]_{\max(A_n+s_j, \bar{A}_m+\bar{s}_k)}^{\min(B_n, \bar{B}_m)}$$
$$+ \left(H_n(B_n) - H_n(\bar{B}_m)\right) I_{(0,\infty)}(B_n - \bar{B}_m) \quad (6)$$

where ${}_2F_1$ is hypergeometric function. A macron is used to explicitly indicate variables related to the second route, $R_2$. $\xi_n$ and $\bar{\xi}_m$ are defined as follows

$$\xi_n \triangleq \frac{1}{(n-1)! \prod_{i=1}^{n}(b_i - a_i)} \quad (7)$$

$$\bar{\xi}_m \triangleq \frac{1}{(m)! \prod_{i=1}^{m}(\bar{b}_i - \bar{a}_i)} \quad (8)$$

**Proof:** Delay inversion occurs whenever the sum of n uniform random variables of the first route is greater than the sum of m uniform random variables of the second route, that is

$$P(R_1 > R_2) = \int_0^\infty \int_0^x h_n(x) \bar{h}_m(y) dy\, dx$$
$$= \int_0^\infty h_n(x) \bar{H}_m(x) dx \quad (9)$$

Since the maximum value of $r_n(x)$ is $2^n$, (4) can be written as follows

$$h_n(x) = \xi_n \sum_{j=1}^{2^n} (-1)^{\|w_{nj}\|} (x - A_n - s_j)^{n-1} I_{(A_n,B_n)}(x) I_{(0,r_n(x))}(j) \quad (10)$$

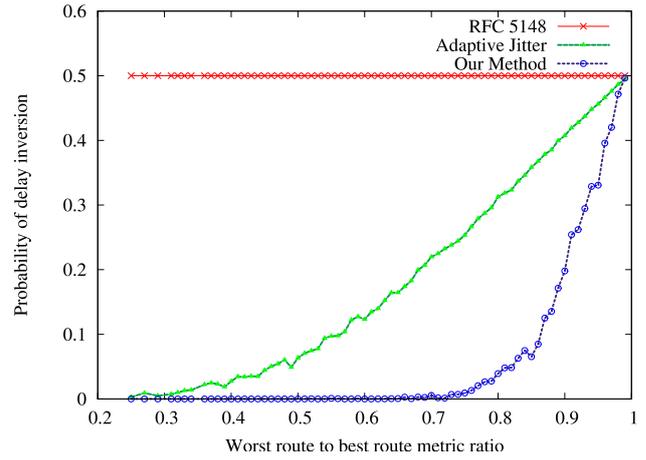

Fig. 2 Probability of delay inversion when two paths have the same length

Hence,

$$P(R_1 > R_2)$$
$$= \xi_n \bar{\xi}_m \int_0^{\bar{B}_m} \sum_{j=1}^{2^n} \sum_{k=1}^{2^m} (-1)^{\|w_{nj}\| + \|\bar{w}_{mk}\|} \frac{(x - A_n - s_j)^{n-1}}{(x - \bar{A}_m - \bar{s}_k)^{-m}}$$
$$\times I_{(A_n,B_n)}(x) I_{(0,r_n(x))}(j) I_{(\bar{A}_m,\bar{B}_m)}(x) I_{(0,\bar{r}_m(x))}(k) dx$$
$$+ \xi_n \int_0^\infty \sum_{j=1}^{2^n} (-1)^{\|w_{nj}\|} (x - A_n - s_j)^{n-1} I_{(A_n,B_n)}(x) I_{(\bar{B}_m,\infty)}(x) I_{(0,r_n(x))}(j) dx \quad (11)$$

Therefore,

$$P(R_1 > R_2)$$
$$= \xi_n \bar{\xi}_m \sum_{j=1}^{2^n} \sum_{k=1}^{2^m} (-1)^{\|w_{nj}\| + \|\bar{w}_{mk}\|} \int_{\max(A_n,\bar{A}_m)}^{\min(B_n,\bar{B}_m)} \frac{(x - A_n - s_j)^{n-1}}{(x - \bar{A}_m - \bar{s}_k)^{-m}}$$
$$\times I_{(0,r_n(x))}(j) I_{(0,\bar{r}_m(x))}(k) dx + I_{(0,\infty)}(B_n - \bar{B}_m)$$
$$\times \int_{\bar{B}_m}^{B_n} \xi_n \sum_{j=1}^{2^n} (-1)^{\|w_{nj}\|} (x - A_n - s_j)^{n-1} I_{(0,r_n(x))}(j) dx \quad (12)$$

For each term in the summations, j and k are constant. So, in order for terms containing $I_{(0,r_n(x))}(j)$ to be non-zero, the value of $r_n(x)$ must be greater than j. Therefore, the smallest value of x that satisfies the condition would be $r_n^{-1}(j) = A_n + s_j$ and consequently $P(R_1 > R_2)$ can be rewriten as follows

$$P(R_1 > R_2) = \xi_n \bar{\xi}_m \sum_{j=1}^{2^n} \sum_{k=1}^{2^m} (-1)^{\|w_{nj}\| + \|\bar{w}_{mk}\|}$$
$$\times \int_{\max(A_n+s_j, \bar{A}_m+\bar{s}_k)}^{\min(B_n,\bar{B}_m)} \frac{(x - A_n - s_j)^{n-1}}{(x - \bar{A}_m - \bar{s}_k)^{-m}} dx$$
$$+ \left(H_n(B_n) - H_n(\bar{B}_m)\right) I_{(0,\infty)}(B_n - \bar{B}_m) \quad (13)$$

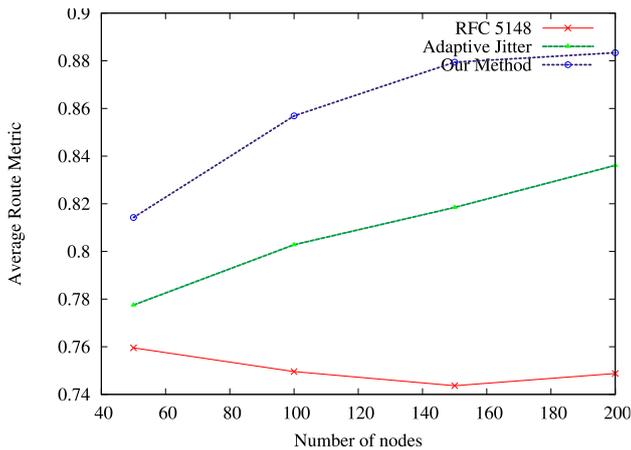

Fig. 3 Average metric of all routes

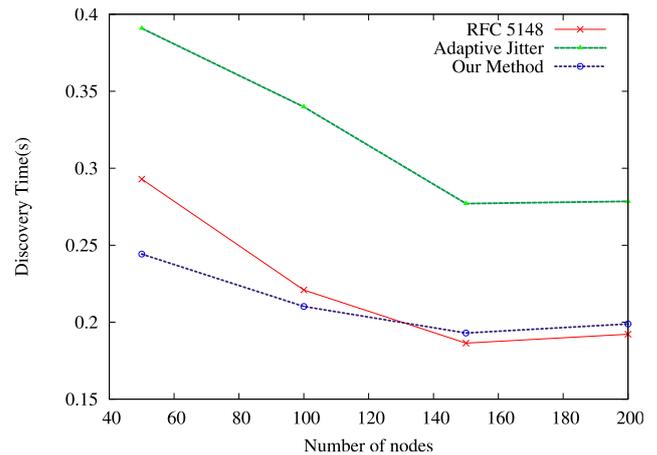

Fig. 4 Average route discovery time

The proof follows immediately by integrating the lat equation.

□

Using theorem 1, we computed average probability of delay inversion when the two paths have the same length, in this case 6 hops, using Maple software. We assumed link metrics are in range (0,1) and route metric is the average metric of its links. We also assumed $J_{max}$ 100ms and C is 30ms. Therefore, the RFC 5148 [1] obtained delay values from $U \sim [0, 100]$, adaptive jitter method from $U \sim [(1 - m)100, 100]$, and our method from $U \sim [(1 - m)100, (1 - m)100 + 30]$. The average probability of delay inversion was obtained by putting over thousand different possible values in equation (6)

As it is shown in Fig 2, hop-based jitter mechanism which ignores link metrics leads to a constant value of 0.5, despite the difference in quality of routes. Our method significantly reduces the probability of delay inversion, even when the two routes' metric are relatively close.

### III. SIMULATIONS

In this section, we present results of a set of simulations conducted by ns-2 to comprehensively show the effect of jitter mechanism on routing. In our simulation, 100 nodes were randomly deployed in a square region of 1000×1000 meters. Transmission range was 250m and $J_m$ was 250ms. Link metrics were selected within $[0.5,1]$ and C was 40ms. The simulations lasted for 100 seconds and 10 random nodes initiated route discovery every 2 seconds. We compared three different mechanisms, that is RFC 5148 ($U \sim [0,250]$), adaptive jitter ($U \sim [(1 - m)250, 250]$) and our method ($U \sim [(1 - m)250, (1 - m)250 + 40]$).

Fig. 3 demonstrates the effect of increasing number of nodes on route metric. As it is expected, our method finds better routes. Additionally, since more nodes mean more possible routes from a source to a destination, routes' metrics increase as number of nodes increases, if the mechanism exploits the opportunity. A mechanism that has the least value of probability of delay inversion is more likely to find optimal routes. That is the reason our method outperforms other methods.

Range and lower bounds of a uniform random variable used to obtain delay in route discovery process can dramatically impact route discovery time. RFC 5148 always uses a fixed range ($U \sim [0,250]$). Adaptive jitter method imposes more average delay than RFC 5148 on low quality links. Moreover, since average forwarding delay of high quality links in adaptive jitter method is as long as that of FRC 5148 method, in general, route discovery process takes longer by adaptive jitter, as it is shown in Fig. 4. In contrast, our method tries to impose least possible delay on high quality links, by limiting the range of the random variable, so as to reduce route discovery time. Therefore, our method finds better routes comparing to other methods, while almost maintaining the lowest route discovery time.

Fig. 5 shows the number of collisions during route discovery phase with different node density. As possible range of uniform random variable decreases, the probability of collision increases since more nodes are likely to forward RREQ packets at a same time. In our simulation range of uniform random variable of RFC 5148 is equal to $J_{max}$ (250ms), that of our method is C (40ms), and adaptive jitter uses a variable range ($0ms < r < 250ms$). Hence, RFC 5148 has the least number of collisions and our method has the highest. However, the differences in the number of collisions are negligible, as it is shown in Fig. 5.

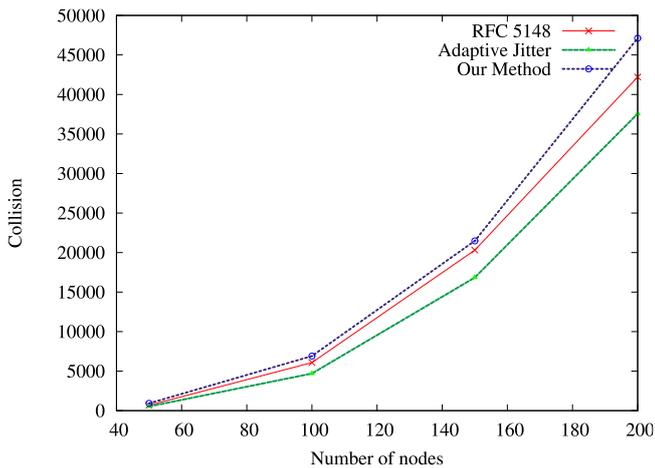

Fig. 5 Number of collision

IV. CONCLUSION

New jitter mechanism based on uniform distribution has been presented in this paper to improve routes found in route discovery stage. Closed form expression has been also derived which allows us to obtain probability of delay inversion of any jitter mechanism using uniform distribution. Using the expression, probability of delay inversion of our mechanism was shown to be lower than other methods. Additionally, ns-2 simulation results have been presented which demonstrate our proposed mechanism provides better and more optimal routes.